\documentclass[aps,prb,twocolumn,superscriptaddress,10pt]{revtex4-2}
\usepackage[justification=centering]{caption}
\usepackage[justification=justified,singlelinecheck=false]{caption}
\usepackage{graphicx}
\usepackage{amsmath}
\usepackage{microtype}
\usepackage{ragged2e}
\usepackage{siunitx}
\usepackage{hyperref}
\usepackage{physics}
\usepackage{caption}
\usepackage{subcaption}
\usepackage{multirow}
\usepackage{xcolor}
\usepackage{lmodern}
\usepackage{parskip}
\usepackage{balance}
\usepackage{graphicx}
\usepackage{caption}
\usepackage{amssymb}
\DeclareUnicodeCharacter{22C5}{\cdot}
\usepackage{subcaption}
\usepackage{float} %
\usepackage{caption}

\begin{document}

\title{Hexagonal Boron Nitride Spin Defects for Quantum Photonics: Annealing-Free Generation by Krypton Ion Implantation}

\author{Ikshvaku Shyam}
\thanks{Equal author contribution}
\affiliation{Laboratory of Optics of Quantum Materials, Department of Physics,
Indian Institute of Technology Bombay, Mumbai 400076, INDIA}

\author{Raj Singh}
\thanks{Equal author contribution}
\affiliation{Department of Chemistry, Indian Institute of Technology Bombay, Mumbai 400076, INDIA}

\author{Mangababu Akkanaboina}
\thanks{Equal author contribution}
\affiliation{Laboratory of Optics of Quantum Materials, Department of Physics,
Indian Institute of Technology Bombay, Mumbai 400076, INDIA}

\author{A. M. Sonawane}
\affiliation{Low Energy Ion Beam Laboratory, Department of Physics,
Savitribai Phule Pune University, Pune 411007, INDIA}

\author{Muthu Satheeshkumar}
\affiliation{Department of Chemistry, Indian Institute of Technology Bombay, Mumbai 400076, INDIA}

\author{Amrita Majumder}
\affiliation{Laboratory of Optics of Quantum Materials, Department of Physics,
Indian Institute of Technology Bombay, Mumbai 400076, INDIA}

\author{Ekta}
\affiliation{Laboratory of Optics of Quantum Materials, Department of Physics,
Indian Institute of Technology Bombay, Mumbai 400076, INDIA}

\author{Janhavi Jayawant Khunte}
\affiliation{Laboratory of Optics of Quantum Materials, Department of Physics,
Indian Institute of Technology Bombay, Mumbai 400076, INDIA}

\author{Akash Khaire}
\affiliation{Low Energy Ion Beam Laboratory, Department of Physics,
Savitribai Phule Pune University, Pune 411007, INDIA}

\author{Kenji Watanabe}
\affiliation{Research Centre for Materials Nanoarchitectonics, National Institute for Materials Science, 1-1 Namiki, Tsukuba 305-0044, JAPAN}

\author{Takashi Taniguchi}
\affiliation{Research Centre for Materials Nanoarchitectonics, National Institute for Materials Science, 1-1 Namiki, Tsukuba 305-0044, JAPAN}

\author{Gopalan Rajaraman}
\email{rajaraman@chem.iitb.ac.in}
\affiliation{Department of Chemistry, Indian Institute of Technology Bombay, Mumbai 400076, INDIA}

\author{S. S. Dahiwale}
\email{ssd@physics.unipune.ac.in}
\affiliation{Low Energy Ion Beam Laboratory, Department of Physics,
Savitribai Phule Pune University, Pune 411007, INDIA}

\author{Anshuman Kumar}
\email{anshuman.kumar@iitb.ac.in}
\affiliation{Laboratory of Optics of Quantum Materials, Department of Physics,
Indian Institute of Technology Bombay, Mumbai 400076, INDIA}
\affiliation{Centre of Excellence in Quantum Information, Computation, Science and Technology, Indian Institute of Technology Bombay, Powai, Mumbai 400076, India}

\begin{abstract}
Controlled, reproducible generation of luminescent defect centres in hBN remains a key challenge for scalable quantum-photonic technologies. Here, we report Kr$^{+}$ ion implantation as a tunable, annealing-free, and chemically inert route to room-temperature near-infrared luminescent spin defects in hBN, requiring no pre- or post-implantation annealing. SRIM Monte Carlo simulations
were used to optimise the parameters for 40~keV Kr$^{+}$ irradiation of hBN flakes. The implanted samples exhibit a stable near-infrared photoluminescence (PL) band centred at $\sim$830~nm whose intensity increases with implantation
fluence over $10^{11}$--$10^{15}$~ions\,cm$^{-2}$. Temperature-dependent PL measurements (20--=300~K) reveal a linewidth broadening well described by a $T^{3}$ dependence, consistent with acoustic-phonon-mediated dephasing. Raman spectra show the characteristic $E_{2g}$ mode of pristine hBN at $\sim$1366~cm$^{-1}$ alongside an implantation-induced defect feature at $\sim$1295~cm$^{-1}$, confirming irradiation-induced lattice disorder. Electron paramagnetic resonance (EPR) measurements reveal a paramagnetic centre with a
$g$-factor of 2.003, and density functional theory (DFT) calculations indicate that a spatially separated $V_{\mathrm{N}}$--$C_{\mathrm{B}}$ donor--acceptor
pair complex is a viable origin of the observed optical and magnetic signatures. Overall, Kr$^{+}$ implantation offers an effective, annealing-free, and scalable
platform for generating stable room-temperature luminescent defects, providing a promising route toward quantum photonics.
\end{abstract}

\maketitle
\section{Introduction}

Luminescent defect centres are promising platforms for quantum technologies, where defect-induced electronic states within the bandgap enable efficient radiative recombination and light emission.\cite{guo2024quantum,aharonovich2016solid} Among solid-state quantum emitters, defect centres in hexagonal boron nitride (hBN) have attracted considerable attention owing to their room-temperature operation, high photostability, and compatibility with quantum communication and integrated photonic platforms.\cite{Singh2026NF,gao2023atomically,vogl2025defects,toth2019single}

Single-photon emitters (SPEs) in two-dimensional materials are particularly attractive because their van der Waals (vdW) nature facilitates integration with photonic architectures through heterostructure engineering.\cite{shaik2021optical} While transition-metal dichalcogenides (TMDCs) such as WSe$_2$ can host single-photon emitters, their operation generally requires cryogenic temperatures, and similarly, defect centres in silicon carbide and III–V quantum dots are susceptible to phonon-induced broadening and environmental fluctuations that often necessitate cryogenic cooling to achieve narrow spectral linewidths.\cite{parto2021defect} In diamond NV centres, strong electron--phonon coupling results in a low Debye-Waller factor ($\sim 3$--$5\%$), causing most emission to be transferred into phonon sidebands rather than the zero-phonon line (ZPL).\cite{kamada2008broadening,plakhotnik2015electron,singh2025plasmonic,jelezko2006single,pezzagna2011creation,chen2022defect}

Hexagonal boron nitride (hBN) addresses these limitations through its wide (indirect and direct) bandgap ($\sim5.9$--$6.1~\mathrm{eV}$), which supports deep, localised defect states that remain electronically isolated from the host bands. This isolation reduces thermal quenching and facilitates room-temperature single-photon emission.\cite{cassabois2016hexagonal} Furthermore, several hBN defect centres exhibit comparatively large Debye--Waller factors ($\sim0.3$--$0.8$), resulting in stronger zero-phonon emission and improved photon coherence. The thin hBN layers facilitate integration with photonic cavities and waveguides.\cite{ccakan2025quantum,tran2016quantum}

A wide range of techniques has been explored to create luminescent defect centres in hBN, including thermal annealing, plasma and chemical etching, focused laser irradiation, strain engineering, electron-beam irradiation, nano-indentation, and ion implantation.\cite{Proscia2018_Strain,Schaumburg2025_PlasmaDefects,Chen2024_WaferScaleSPE,Guo2022_SpinDefects,Huang2022_Implantation,ccakan2025quantum,majumder2026deterministicsinglephotonsourceshexagonal,akkanaboina2026deterministicsinglephotonemitterarrays,Singh2025ami} Among these approaches, ion implantation offers control over defect density, depth, and damage through tuning of ion mass, energy, and fluence.\cite{Guo2022_SpinDefects,Huang2022_Implantation,Venturi2024_LPR}

The optical response of ion-implanted hBN depends on the implanted species and implantation conditions, which govern vacancy production, defect-complex formation, and the resulting electronic structure of the host lattice. Although ion implantation is an effective method for engineering colour centres in hBN, the microscopic origin of many near-infrared emitters remains unclear.\cite{Petrov2024_HeIrrad,Huang2022_Implantation,Reimers2020_VBPhotophysics,Guo2022_SpinDefects} Luminescent centres in the 800--850~nm range have been observed following irradiation or implantation, yet definitive identification of the underlying defect configurations is complicated by the coexistence of multiple vacancy complexes, charge states, and impurity-related defects.\cite{Reimers2020_VBPhotophysics,Guo2022_SpinDefects}

Among the noble-gas ions investigated, the number of vacancies, displacements, and replacement collisions increases rapidly from He$^{+}$ to Kr$^{+}$ owing to the progressively larger momentum transfer and enhanced nuclear stopping associated with increasing ion mass. Beyond Kr$^{+}$, however, both the vacancy yield and displacement production approach saturation, with Xe$^{+}$ and Rn$^{+}$ providing only marginal increases in damage per ion despite their substantially larger masses. At the same implantation energy, heavier ions also exhibit a reduced projected range and are confined to a shallower implantation volume, limiting the spatial extent of defect generation. In addition, the ionisation potential decreases systematically from He to Kr and changes only weakly for heavier noble gases. Consequently, further increases in atomic mass beyond Kr do not translate into a proportionate enhancement of defect production. Kr$^{+}$ therefore represents an optimal balance between damage generation and implantation depth, producing nearly the maximum vacancy and displacement density achievable within the noble-gas series while maintaining a sufficiently large projected range for volumetric defect engineering. These factors motivated the selection of Kr$^{+}$ as the implantation species for controlled clean, inert colour-centre generation in hBN.

Electron paramagnetic resonance (EPR) spectroscopy directly probes implantation-induced spin defects and, when combined with first-principles calculations, enables their identification at the microscopic level. This is particularly relevant for donor–acceptor-pair (DAP) defects, where recombination between spatially separated donor and acceptor states can yield stable near-infrared luminescence with retained paramagnetism.

We report spin defects in hBN generated by 40~keV Kr$^{+}$ implantation, which produce a near-infrared emission band centred at $\sim$830~nm without post-implantation annealing. Temperature-dependent photoluminescence (PL), electron paramagnetic resonance (EPR), and density functional theory (DFT) calculations were used to investigate the defect origin and spin properties. The emission remains stable from 20–300 K, with linewidth broadening attributed to acoustic-phonon coupling. EPR measurements reveal the formation of spin-active defect centres, while first-principles calculations suggest that nitrogen-vacancy-related donor-acceptor-pair complexes are energetically favourable candidates for the observed optical and magnetic signatures. These findings establish Kr$^{+}$ implantation as an effective route for creating stable near-infrared spin defects in hBN and provide insight into the defect configurations responsible for their optical and spin properties.

\begin{figure*}[t]
\centering
\includegraphics[width=\linewidth]{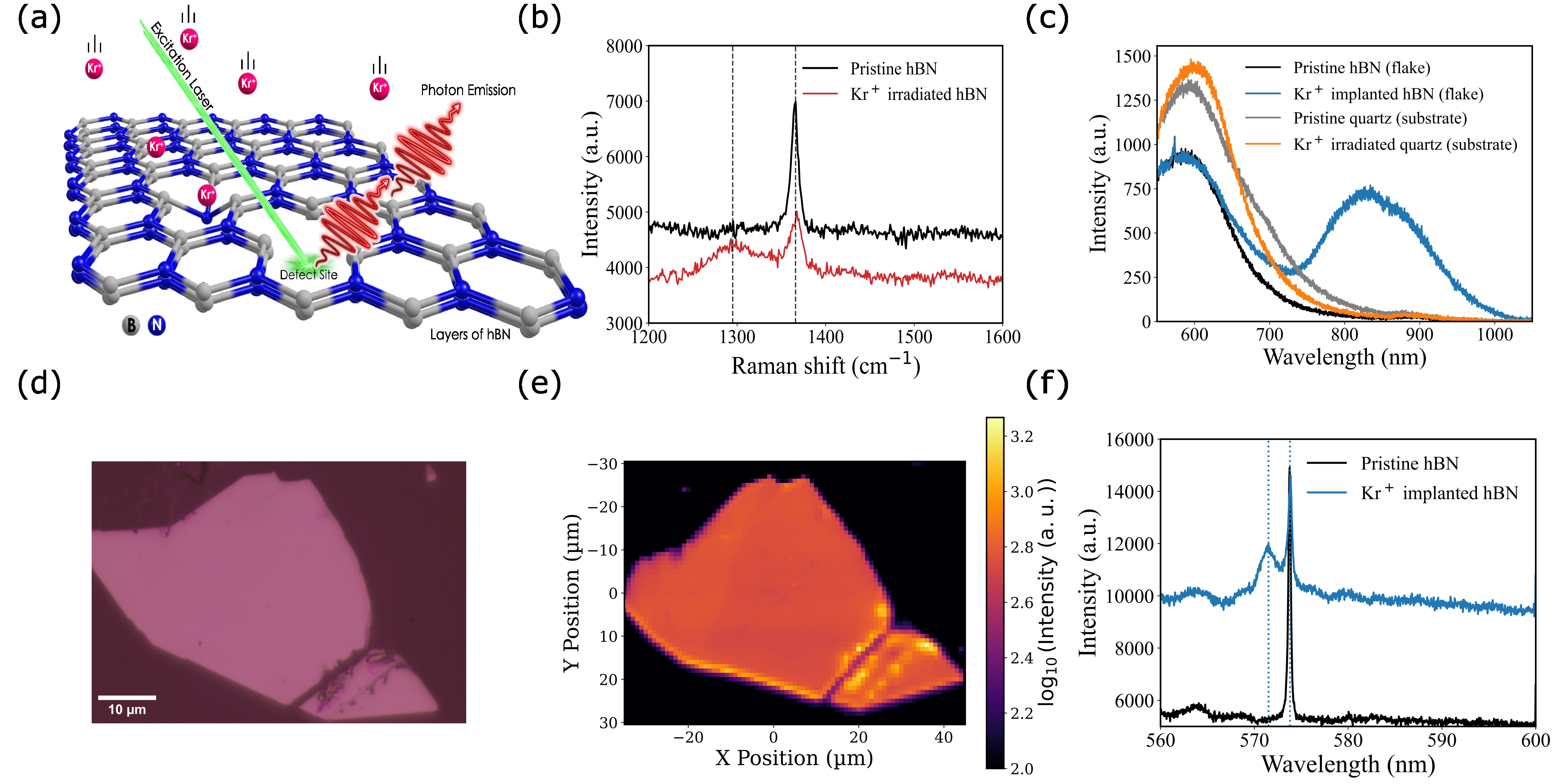}
\caption{\justifying
\textbf{Optical characterization of Kr$^{+}$-irradiated hBN.}
(a) Schematic illustration of Kr$^{+}$ ion implantation and optical excitation using a 532~nm green laser in few-layer hBN for defect-related near-IR emission.
(b) Raman spectra of pristine and Kr$^{+}$-irradiated hBN showing $E_{2g}$ and defect modes.
(c) PL spectra of pristine and Kr$^{+}$-implanted hBN flakes along with the quartz substrate backgrounds.
(d) Optical image of an exfoliated hBN flake on quartz (scale bar: $10~\mu\mathrm{m}$).
(e) Spatial PL intensity map (at 830 nm) of the hBN flake showing the uniform defect PL distribution.
(f) PL spectrum in the $560$--$600~\mathrm{nm}$ range showing Raman scattering peaks corresponding to the $\sim 1295~\mathrm{cm^{-1}}$ and $\sim 1366~\mathrm{cm^{-1}}$ vibrational modes at 571.3 nm and 573.7 nm respectively, when excited by a 532 nm laser.}
\label{fig:fig1}
\end{figure*}

\section{Results and Discussion}
\subsection{Optical Characterization of Kr$^{+}$-Implanted hBN}

To minimise irradiation-induced damage to the underlying substrate, hBN flakes approximately 100 nm thick were selected for implantation. Prior to irradiation, Monte Carlo simulations were performed using the full-cascade damage module of the SRIM package to investigate the interaction of 40~keV Kr$^{+}$ ions with hBN, to estimate ion penetration depths, and determine suitable implantation conditions for defect generation (see Section S3, Supplementary Information).\cite{ziegler2010srim} The simulations predicted a projected end-of-range (EOR) of approximately 30.5~nm, ensuring that the implanted ions and the associated damage profile remained confined within the hBN layer. As SRIM is based on the binary collision approximation (BCA), the predicted damage profile should be regarded as an approximation for the anisotropic layered structure of hBN.\cite{nordlund2018primary}

Unlike many previous implantation studies that require post-implantation annealing at temperatures exceeding 800 °C, the present defects become optically active immediately after implantation, suggesting either direct formation of radiative defect complexes or efficient room-temperature defect reconstruction.

Figure~1(a) schematically illustrates the generation of luminescent defect centres in hBN by Kr$^{+}$ ion implantation. Upon implantation at normal incidence, 40~keV Kr$^{+}$ ions transfer energy to the lattice through nuclear and electronic stopping, displacing boron and nitrogen atoms and creating vacancies and vacancy-related defect complexes. These defects introduce localised electronic states within the wide bandgap of hBN.\cite{abdi2018color,weston2018native} Under 532~nm optical excitation, the defect centres act as radiative recombination sites, producing the observed photoluminescence emission.\cite{toth2019single} 

\begin{figure*}[t]
\centering
\includegraphics[width=0.85\linewidth]{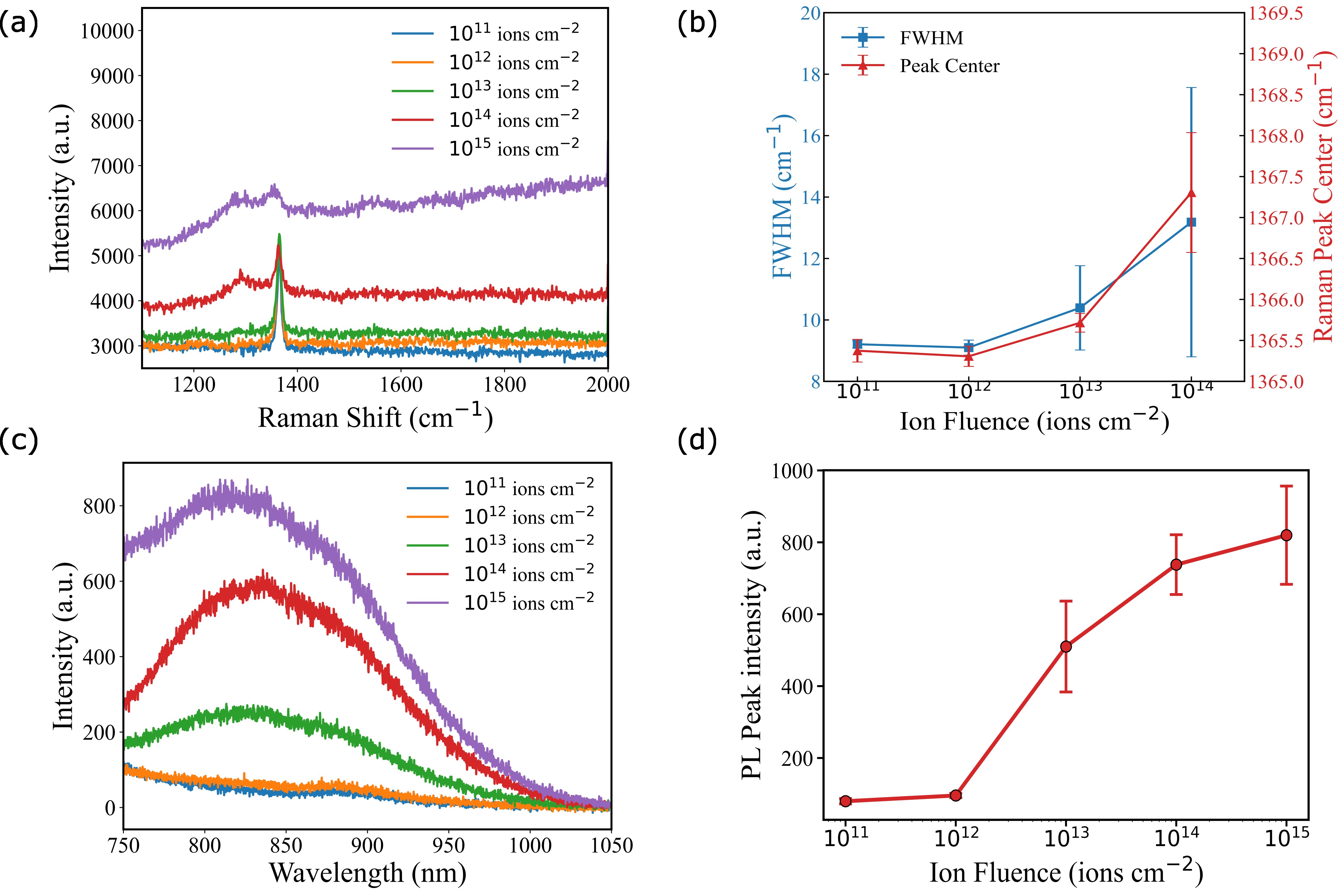}
\caption{\justifying
\textbf{Fluence-dependence of Kr$^{+}$-implanted hBN.}
(a) Raman spectra of hBN for different Kr$^{+}$ ion fluences showing the evolution of Raman mode near $\sim1295~\mathrm{cm^{-1}}$.
(b) Raman full width at half maximum (FWHM) and peak position variation as a function of implantation fluence.
(c) Defect-related PL emission with an increase in implantation fluence.
(d) The PL peak intensity variation with increase in implantation fluence}
\label{fig:fig2}
\end{figure*}

The Raman spectra of pristine and Kr$^{+}$-irradiated hBN are shown in Figure~1(b), acquired under identical excitation conditions. Pristine hBN exhibits the characteristic $E_{2g}$ in-plane phonon mode at $\sim1366~\mathrm{cm^{-1}}$, consistent with an ordered sp$^{2}$-bonded lattice.\cite{Kim2023_ColorCenters,stenger2017low} Following Kr$^{+}$ implantation, an additional Raman feature emerges near $\sim1295~\mathrm{cm^{-1}}$, accompanied by broadening and reduced intensity of the $E_{2g}$ mode. The appearance of this feature is consistent with implantation-induced disorder and lattice distortion. Ion-implantation-induced defects break translational symmetry and relax Raman selection rules, enabling finite-wavevector phonons ($\mathbf{q}\neq0$) to contribute to the spectrum beyond the zone-centre modes ($\mathbf{q}\approx0$). Consequently, disorder-activated vibrational modes associated with the phonon density of states become Raman-active, giving rise to the observed feature near $\sim1295~\mathrm{cm^{-1}}$, which thus serves as a Raman fingerprint of implantation-induced lattice disorder in hBN.

The PL spectra of pristine hBN (black curve) and Kr$^{+}$-implanted hBN flakes (blue curve) on the quartz substrate (pristine quartz and implanted quartz, grey and orange curves, respectively) are compared in Figure~1(c). The quartz substrate exhibits a weak, broadband emission centred around ~620 nm (~2.0 eV), which is attributed to defect-related luminescence in SiO$_2$, most likely arising from non-bridging oxygen hole centres (NBOHCs). \cite{skuja2020luminescence,skuja1998optically,pacchioni1997computed}This emission is also partially visible in the pristine hBN flake spectrum due to the thin nature of exfoliated flakes and the contribution of substrate luminescence through the flake. The absence of any PL peak around $830~\mathrm{nm}$ in the non-irradiated (pristine) hBN suggests that our samples are clean, without intrinsic luminescent point defects in that region. In contrast, the irradiated regions of the sample exhibit emission peaks centred near $\sim 830$~nm (1.5 eV) at room temperature. The quartz substrate without hBN was irradiated using the same irradiation conditions. The absence of emission near 830 nm in irradiated quartz suggests that the observed PL primarily originates from the implanted hBN flakes. 

Figure~1(d) shows an optical image of an hBN flake irradiated with a fluence of $10^{14}$~ions~cm$^{-2}$. No discernible change in the optical appearance of the flake is observed after irradiation, indicating that Kr$^{+}$ implantation does not produce visible damage under the present conditions. Figure~1(e) presents a spatial PL intensity map of the defect emission centred near 830~nm, acquired from the same flake using a 532~nm excitation laser at 11.1 mW with a step size of 1~$\mu$m. The defect-related emission is observed across the irradiated region, indicating a relatively uniform distribution of optically active implantation-induced defects. Enhanced emission near flake edges likely reflects local variations in strain or defect density. Pristine hBN shows no PL at 830 nm; its emergence post-implantation confirms the emission originates from Kr$^{+}$-irradiated defect centres.\cite{nai2015studying}

The Raman spectrum in the 560--600~nm spectral window, acquired using a 2400~lines/mm grating. The measurement resolves the characteristic hBN $E_{2g}$ phonon mode at 573.7~nm ($1366~\mathrm{cm^{-1}}$) together with a disorder-activated Raman mode at 571.3~nm ($1295~\mathrm{cm^{-1}}$) as distinct spectral features are shown in Figure~1(f). The PL excitation to detect these fine peaks was performed with a 2400 lines/mm grating, a 5-second exposure, 20 mW of incident laser power, and 2 accumulations to observe the fine, close-by peaks. Two peaks are observed at 573.7~nm and 571.3~nm, corresponding to Raman shifts of 1366~cm$^{-1}$ and 1295~cm$^{-1}$, respectively, relative to the 532~nm excitation wavelength. The feature at 573.7~nm is present in both pristine and Kr$^{+}$-implanted hBN and is assigned to the characteristic $E_{2g}$ phonon mode of hBN. In contrast, the 571.3~nm feature appears exclusively in the implanted samples and is attributed to the disorder-activated Raman mode identified in Figure~1(b). Its emergence provides further evidence of implantation-induced lattice disorder and corroborates the successful generation of irradiation-induced defects in hBN. We further study the vibrational and optical properties as a function of implantation fluence.

The Raman spectra variation of hBN flakes implanted with Kr$^{+}$ ions at different ion fluences ranging from $10^{11}$ to $10^{15}$~ions~cm$^{-2}$ is shown Figure~2(a).The characteristic $E_{2g}$ in-plane phonon mode near 1366~cm$^{-1}$ remains clearly visible up to a fluence of $10^{14}~\mathrm{ions~cm^{-2}}$, indicating that the hexagonal lattice retains a significant degree of long-range order within this fluence range. In contrast, at $10^{15}~\mathrm{ions~cm^{-2}}$, the $E_{2g}$ mode is no longer resolved, and the spectrum becomes dominated by a broad defect-related luminescence background. This behaviour suggests that the accumulated irradiation damage at the highest fluence approaches or exceeds the amorphisation threshold of hBN.

Raman spectra were fitted with Voigt profiles to extract the $E_{2g}$ peak position and FWHM, both of which are plotted as a function of implantation fluence in Figure~2(b). At $10^{11}$--$10^{12}$~ions~cm$^{-2}$, the peak position ($\sim1365.6~\mathrm{cm^{-1}}$) and linewidth ($\sim9~\mathrm{cm^{-1}}$) remain comparable to those of pristine hBN within experimental uncertainty, indicating that structural modifications at these fluences are below the Raman detection threshold. As the fluence increases to $10^{13}$ and $10^{14}$~ions~cm$^{-2}$, two systematic trends emerge. First, the FWHM increases from $\sim9~\mathrm{cm^{-1}}$ to $\sim13~\mathrm{cm^{-1}}$, consistent with enhanced phonon scattering and reduced phonon coherence resulting from the accumulation of vacancies and defect complexes.\cite{cassabois2021hexagonal,stenger2017low} Second, the $E_{2g}$ peak exhibits a gradual blueshift of approximately $1.5~\mathrm{cm^{-1}}$.\cite{kirchhoff2022electronic} Although vacancy formation is often associated with phonon softening, the observed blueshift may arise from irradiation-induced compressive strain in the substrate-supported flakes, potentially combined with phonon-confinement effects within the remaining ordered crystalline regions. 

The $10^{15}$~ions~cm$^{-2}$ sample was excluded from the peak-position and FWHM analysis because the characteristic $E_{2g}$ Raman mode could no longer be reliably resolved or fitted with a Voigt profile. Instead, the spectrum is dominated by a broad defect-related luminescence background, indicating extensive irradiation-induced disorder and severe degradation of the crystalline hBN lattice. 

\begin{figure*}[t]
\centering
\includegraphics[width=0.9\linewidth]{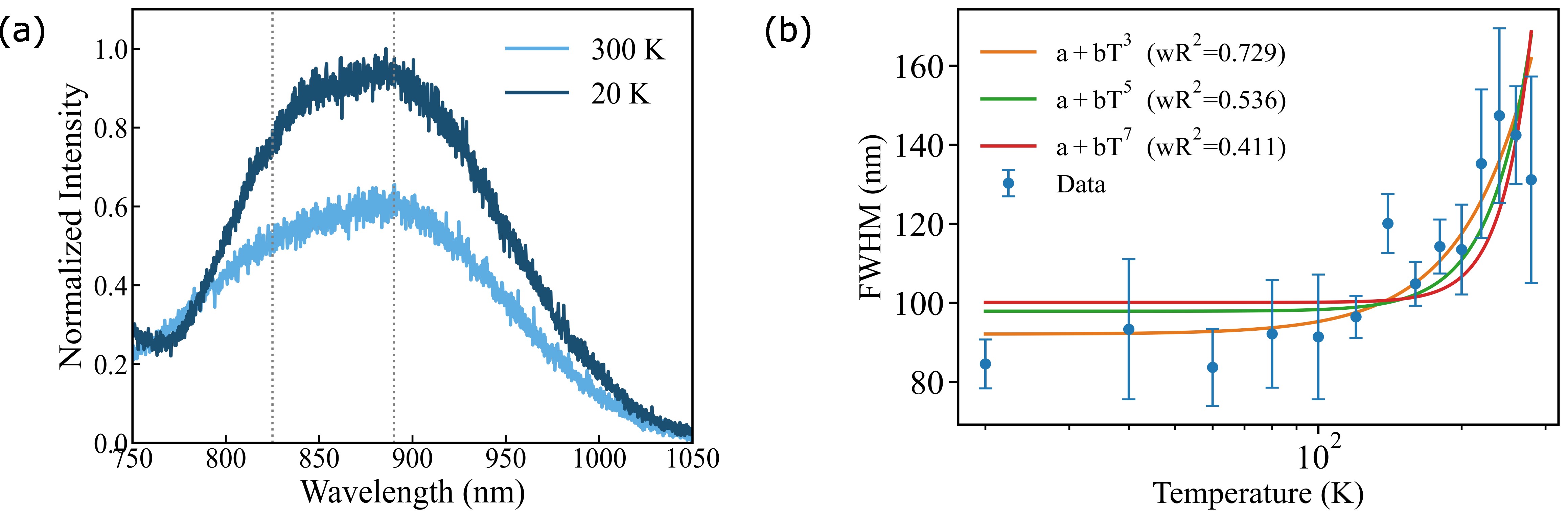}
\caption{\justifying
\textbf{Temperature-dependent PL characteristics of Kr$^{+}$-implanted hBN.}
(a) PL spectra at 20~K and 300~K, showing stable defect-emission intensity at low temperature. 
(b) Temperature dependence of the defect-emission full width at half maximum (FWHM), fitted using $a+bT^{3}$, $a+bT^{5}$, and $a+bT^{7}$ models to describe phonon-induced broadening.}
    \label{fig:fig3}
\end{figure*}

To contextualise the observed $\sim1.5~\mathrm{cm^{-1}}$ blueshift within an established physical framework, the phonon-frequency shift can be discussed using the Grüneisen formalism.\cite{cuffari2020calculation} For the in-plane $E_{2g}$ phonon mode, the frequency shift under biaxial strain $\varepsilon$ is given by

\begin{equation}
\frac{\Delta\omega}{\omega_0} = -2\gamma\varepsilon,
\end{equation}

where $\gamma$ is the mode Grüneisen parameter, $\omega_0$ is the unstrained phonon frequency, and $\Delta\omega$ is the strain-induced frequency shift. Previous studies have reported a negative value of $\gamma$ for the $E_{2g}$ mode of hBN,\cite{blundo2022vibrational,cai2017raman}, implying that compressive strain ($\varepsilon < 0$) results in a positive frequency shift. The observed blueshift is therefore consistent with irradiation-induced modifications of the local bonding environment and may also reflect the presence of compressive residual strain generated during implantation.\cite{seo2025damage}

Figure~2(c) shows the evolution of the PL spectra with implantation fluence. At low fluences,  $10^{11}$ and $10^{12}~\mathrm{ions~cm^{-2}}$, the near-infrared emission is extremely weak, indicating a low density of luminescent defect centres. As fluence increases, a broad emission band centred between $\sim820$ and $850$~nm emerges and increases systematically in intensity under identical measurement conditions. This behaviour is consistent with the progressive formation of implantation-induced defect centres that introduce localised electronic states within the hBN bandgap and serve as radiative recombination centres. The monotonic increase in PL intensity therefore provides direct evidence for the controlled creation of luminescent defect centres by Kr$^{+}$ irradiation. Notably, measurable PL emission persists at a fluence of $10^{15}~\mathrm{ions~cm^{-2}}$ despite the disappearance of a well-defined crystalline $E_{2g}$ Raman signature, indicating that luminescent defect centres remain optically active even under conditions of severe irradiation-induced lattice disorder and loss of long-range crystalline order.

The PL intensity of the $\sim$830~nm band as a function of implantation fluence is shown in figure~2(d). Emission remains weak and nearly unchanged at $10^{11}$--$10^{12}$~ions~cm$^{-2}$, indicating a low density of optically active centres. A pronounced increase occurs between $10^{12}$ and $10^{13}$~ions~cm$^{-2}$, consistent with rapid growth in luminescent defect concentration. At higher fluences ($10^{14}$--$10^{15}$~ions~cm$^{-2}$), the intensity starts to saturate.

The PL spectra reported here originate from an ensemble of implantation-induced defect centres distributed throughout the irradiated hBN flakes; consequently, the broad emission band centred at $\sim830$~nm reflects the ensemble-averaged optical response rather than that of an individual emitter. This interpretation is supported by second-order photon autocorrelation measurements, which yield a correlation function with $g^{(2)}(0)\approx1$ and no observable antibunching dip under 532~nm excitation (see Section~S9, Supplementary Information), confirming the absence of single-photon emission under the present measurement conditions.\cite{loudon1974quantum,kimble1977photon} The ensemble nature of the emission is consistent with the relatively high implantation fluences employed in this work. Accessing the single-emitter regime would likely require substantially lower implantation fluences to produce spatially isolated defect centres.

Additional analyses supporting the implantation strategy and the interpretation of the defect are provided in the Supplementary Information. Section S1 presents energy transfer and bond-breaking by 40 keV Kr$^{+}$ Ions in hBN. Section~S2 outlines how defect-induced symmetry breaking can generate optically active in-gap states in hBN. The thickness of the exfoliated hBN flakes is characterised by AFM in Section~S4, while fluence-dependent PL is shown in Section~S7.

\subsection{Temperature-Dependent and Time-Resolved Optical Measurements}

The Raman and photoluminescence results demonstrate that Kr$^{+}$ implantation provides an effective route for generating luminescent defect centres in hBN. To further probe their optical properties, temperature-dependent photoluminescence measurements were performed over 20-300~K.

Figure~3(a) compares the PL spectra recorded at 300~K and 20~K from Kr$^{+}$-implanted hBN. The low-temperature measurements were carried out in a cryogenic vacuum chamber equipped with quartz optical windows. In addition to the defect-related near-infrared emission, a broad background feature centred near 880 nm was observed, originating from the quartz optical path in the cryogenic setup. At low temperatures, spectral narrowing of the defect emission results in partial overlap with the quartz-related contribution. Therefore, all spectra were analysed using a two-component Voigt deconvolution, yielding a defect-emission peak centred near 830~nm and a broader quartz-related peak near 880~nm, as indicated by the vertical dotted lines in Figure~3(a). All linewidth values discussed below were extracted from the deconvoluted defect-emission component.

The defect-related emission persists throughout the investigated temperature range and exhibits a pronounced reduction in full width at half maximum (FWHM) upon cooling, consistent with suppressed phonon-assisted broadening and reduced optical dephasing. This behaviour confirms the stability of the implantation-induced luminescent centres at cryogenic temperatures.

The temperature dependence of the deconvoluted defect-emission linewidth is presented in Figure~3(b). The linewidth increases with temperature, indicating enhanced coupling of the optical transition to lattice vibrations. The data were fitted using phenomenological expressions of the form \(\Gamma(T)=a+bT^{n}\), with \(n=3\), 5, and 7. The \(T^{3}\) model provided the best description of the experimental data, yielding \(a = 92.11 \pm 3.15~\mathrm{nm}\) and \(b = (3.18 \pm 0.57)\times10^{-6}~\mathrm{nm\,K^{-3}}\), with a weighted coefficient of determination of \(wR^{2}=0.729\) and an Akaike Information Criterion (AIC) value of 15.66. In comparison, the \(T^{5}\) and \(T^{7}\) models yielded lower \(wR^{2}\) values of 0.536 and 0.411, respectively, together with higher AIC values of 23.96 and 29.30. These results indicate that the \(T^{3}\) dependence provides the most appropriate empirical description of the temperature evolution of the linewidth over the investigated temperature range.\cite{hazra2026temperature} Although the relatively large experimental uncertainties-primarily arising from mechanical vibrations of the cryostat and closed-cycle refrigeration system during data acquisition—preclude a definitive assignment of the underlying broadening mechanism, the observed trend is consistent with increasing electron-phonon coupling and temperature-induced optical dephasing.

\begin{figure*}[t]
\centering
\includegraphics[width=1\linewidth]{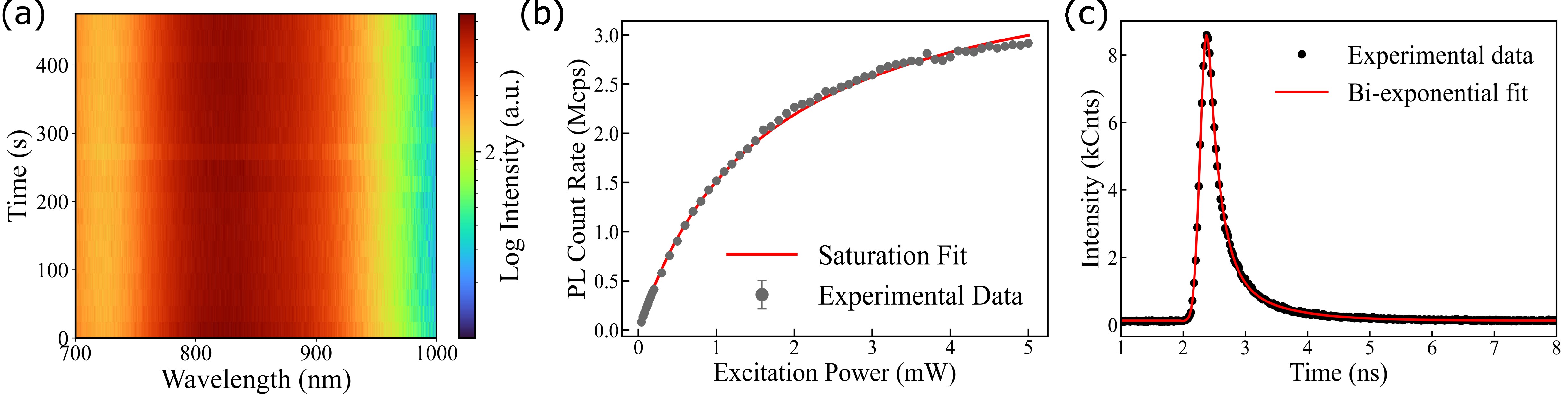}
\caption{\justifying
\textbf{Photostability, power-dependent saturation, and excited-state lifetime characterisation of the Kr$^{+}$-induced defect emitter in hBN.}
(a) Time-dependent PL intensity map recorded over several minutes, demonstrating stable defect emission. (b) Power-dependent PL count rate measured using an Excelitas SPAD. The experimental data (grey circles) are fitted with a standard saturation model (red line), yielding a saturation count rate of \(I_{\mathrm{sat}} = 3.97\) Mcps and a saturation power of \(P_{\mathrm{sat}} = 1.63\) mW.
(c) Defect-emission lifetime measurement. Blue points show the experimental decay, and the orange line represents the fit.}
\label{fig:PL_stability_saturation_TRPL}
\end{figure*}

We note that the relatively large error bars arise from measurement uncertainty associated with low-temperature data acquisition, including fluctuations in signal intensity and spectral fitting due to mechanical vibrations and thermal instabilities of the cryogenic measurement system during chiller operation.

Figure~4(b) shows the excitation-power dependence of the defect PL intensity under 532~nm excitation, measured using an Excelitas SPAD detector. A 590 nm long-pass filter was placed in the collection path to suppress residual excitation laser light and first-order Raman scattering from hBN, ensuring that the detected signal originates predominantly from the defect emission. The experimental data were fitted using this model

\begin{equation}
I(P)=I_{\mathrm{sat}}\frac{P}{P+P_{\mathrm{sat}}}+I_{\mathrm{bg}},
\end{equation}

where $I_{\mathrm{sat}}$, $P_{\mathrm{sat}}$, and $I_{\mathrm{bg}}$ denote the saturation intensity, saturation power, and background intensity, respectively.

The fit yields a background count rate of
$I_{\mathrm{bg}} = 0.00 \pm 140.38~\mathrm{counts\,s^{-1}}$, a saturation intensity of $I_{\mathrm{sat}}=(3.97\pm0.01)\times10^{6}~\mathrm{counts\,s^{-1}}$, and a saturation power of $P_{\mathrm{sat}}=1.6255\pm0.0009~\mathrm{mW}$.
The fit describes the data well, with a coefficient of determination of $R^{2}=0.9983$. The observed saturation behaviour shows the emission originates from implantation-induced defect centres. We note that this is the ensemble saturation count rate originating from the collection volume, not a per-emitter brightness; no per-emitter brightness can be inferred without knowledge of the emitter number.

The time-resolved photoluminescence (TRPL) decay measured using time-correlated single-photon counting (TCSPC) is shown in Figure~4(c). Measurements were performed on the $10^{14}$~ions\,cm$^{-2}$ Kr$^{+}$-implanted hBN flakes using a MicroTime~200 lifetime measurement system. Excitation was provided by a 532~nm pulsed laser operating at a repetition rate of 20~MHz, and the emitted photoluminescence was collected through a 532~nm long-pass filter.

The decay curves were analysed using the reconvolution fitting routine implemented in the MicroTime~200 software, which automatically accounts for the instrument response function (IRF). The luminescence decay was modelled using a bi-exponential function ($n=2$),

\begin{equation}
S(t)=\sum_{i=1}^{2} A_i \exp\left(-\frac{t}{\tau_i}\right),
\end{equation}

where $A_i$ and $\tau_i$ represent the amplitudes and lifetimes of the individual decay channels, respectively. The measured signal is obtained through the convolution of the decay model with the IRF,

\begin{equation}
I(t)=\int_{0}^{t}R(\tau)\,S(t-\tau)\,d\tau,
\end{equation}

where $R(t)$ is the instrument response function.

The TCSPC decay is well described by the bi-exponential reconvolution model, yielding amplitudes of $A_1 = 2.00 \pm 0.17$~kCnts and $A_2 = 13.00 \pm 0.13$~kCnts, with corresponding lifetimes of $\tau_1 = 0.80 \pm 0.03$~ns and $\tau_2 = 0.195 \pm 0.007$~ns. The decay is dominated by the fast component ($\tau_2$), which contributes approximately 87\% of the total emission amplitude, indicating that the defect emission is primarily governed by a rapid recombination channel.

\begin{figure*}[t]
\centering
\includegraphics[width=0.85\linewidth]{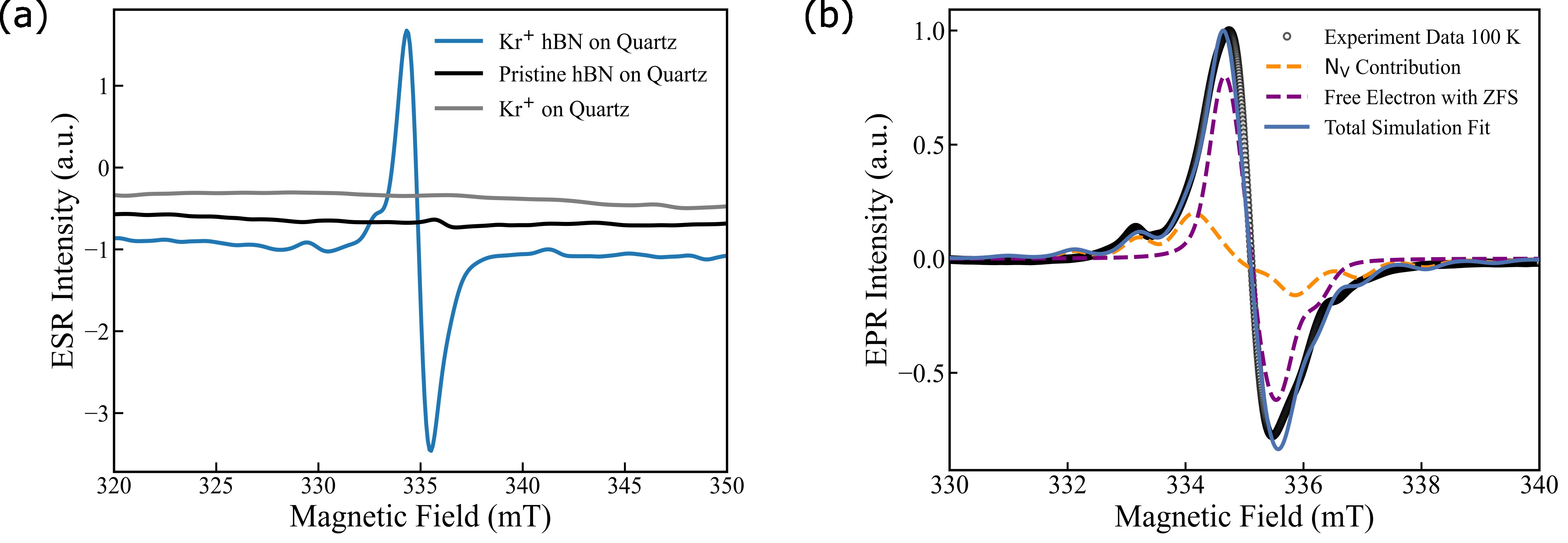}
\caption{\justifying
\textbf{Electron paramagnetic resonance (EPR) characterisation of Kr$^{+}$ defect centres in hBN.}
(a) Room-temperature X-band ESR spectra of Kr$^{+}$-implanted hBN on quartz, pristine hBN on quartz, and Kr$^{+}$-irradiated quartz control substrate. A resonance centred near $\sim3360~\mathrm{G}$ is observed only for Kr$^{+}$-implanted hBN, confirming the formation of irradiation-induced paramagnetic defect centres.
(b) EPR spectrum recorded at $100~\mathrm{K}$ using the same simulation parameters as those used at room temperature.
}
\label{fig:EPR}
\end{figure*}

\subsection{EPR Measurements and First-Principles Defect Modelling}
X-band electron paramagnetic resonance (EPR) spectroscopy was performed to investigate the nature of defects generated in Kr$^{+}$-irradiated hBN. Even at room temperature, the EPR spectrum as shown in Figure 5(a) exhibits a pronounced resonance centred near 335~mT at a microwave frequency of 9.386~GHz, corresponding to an isotropic $g$-value of 2.003. This confirms the presence of paramagnetic species in the irradiated hBN. No comparable signal is observed for pristine hBN or for Kr$^{+}$-irradiated quartz in the absence of an hBN layer, confirming that the EPR signal originates from defects formed within the Kr$^{+}$-irradiated hBN layer. Weak hyperfine features are also observed at room temperature. To improve the hyperfine resolution, EPR measurements were performed at $100~\mathrm{K}$ as shown in Figure 5(b). Although the lower temperature resulted in improved spectral resolution, the hyperfine splitting was not well resolved. Since the sample is a thin film, angle-dependent EPR measurements were also carried out to probe possible orientation-dependent splittings. Although only small variations in the g-value and hyperfine parameters are observed with sample rotation, this indicates that the defect centres are not preferentially aligned and are distributed over multiple orientations.

\begin{figure*}[t]
\centering
\includegraphics[width=\linewidth]{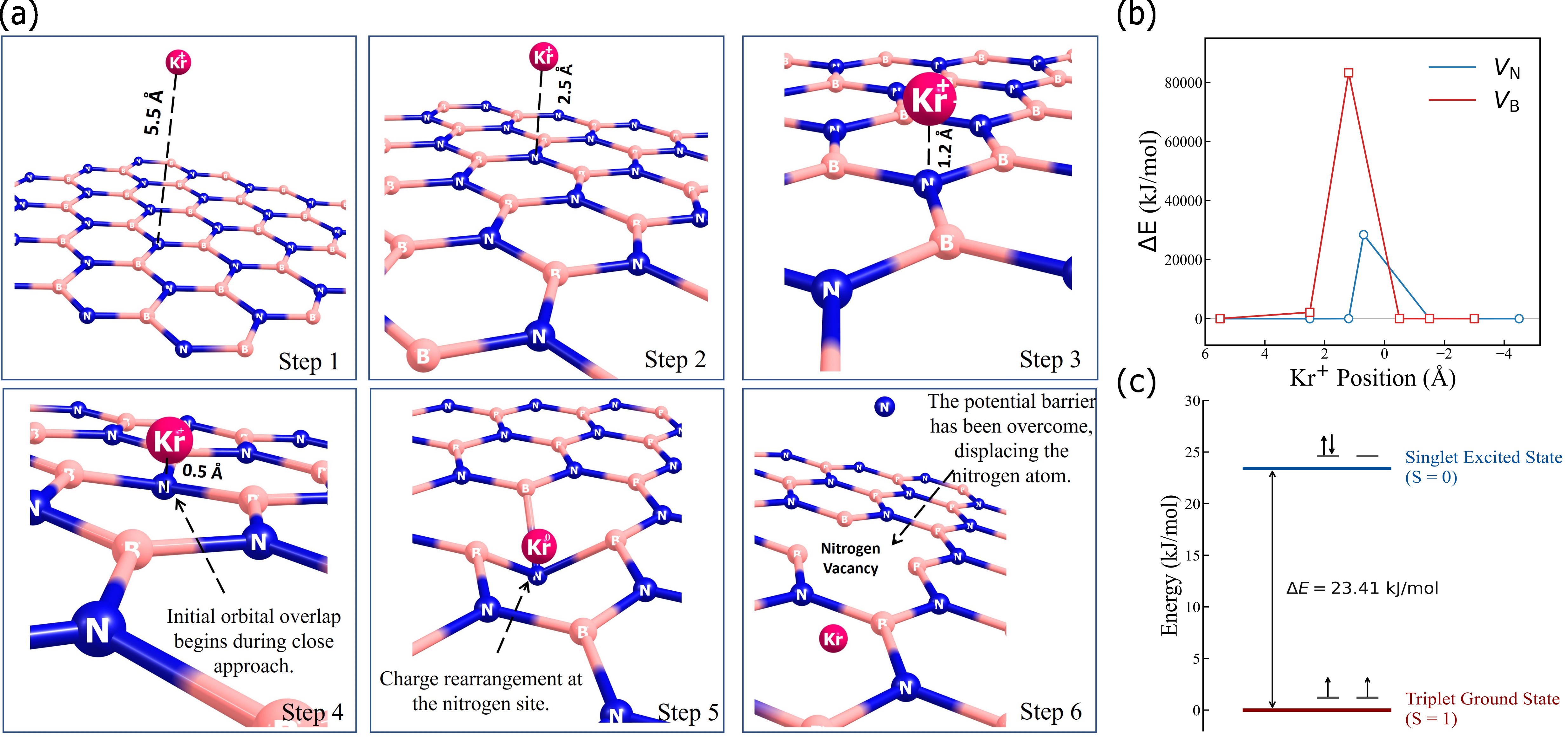}
\caption{\justifying
\textbf{DFT investigation of Kr$^{+}$-induced defect formation and spin-state energetics of $V_{\mathrm{N}}$--$C_{\mathrm{B}}$($\sim 9$~\AA\ separation) defect in hBN.}
(a) Representative configurations along the constrained Kr approach pathway toward the hBN monolayer along the $z$-axis. Individual panels describe the real-space progression: (Steps 1--2) far-separation regime ($+5.5$ \AA\ to $+2.5$ \AA); (Step 3) onset of short-range interaction ($+1.2$ \AA); (Steps 4--5) close-approach region ($+0.5$ \AA\ to the surface plane), and (Step 6) final relaxed configuration showing vacancy formation after geometry relaxation.
(b) Calculated interaction energy profiles ($\Delta E$, kcal mol$^{-1}$) as a function of the Kr$^{+}$ position (\AA) relative to the hBN plane ($z = 0$ \AA) for nitrogen-vacancy ($V_\mathrm{N}$, blue) and boron-vacancy ($V_\mathrm{B}$, red) formation pathways.
(c) Schematic spin-state splitting of the $V_{\mathrm{N}}$--$C_{\mathrm{B}}$ defect in hBN, showing singlet ($S=0$) and triplet ($S=1$) states.}
\label{fig:DFT_defect_energy_spin}
\end{figure*}

EasySpin simulations were performed to explain the origin of the EPR signal and the weakly resolved hyperfine splitting. The experimental spectrum cannot be satisfactorily reproduced by a single spin system, indicating the presence of multiple defect species generated by high-energy Kr$^{+}$ -ion irradiation. The dominant central feature is consistent with defect centres, as expected for a negligible hyperfine interaction, and is consistent with impurity-related defects involving nuclei with zero nuclear spin. In contrast, the weak hyperfine features are consistent with defect centres such as nitrogen or boron vacancies, involving coupling to neighbouring $^{11}$B ($I = 3/2$) or $^{14}$N ($I = 1$) nuclei, respectively. The asymmetric central peak was further examined in terms of either dipolar-coupled spins or $g$-strain arising from a distribution of defect environments.

In the first case, the best fit was obtained using an $S = 1$ component with an anisotropic $g$-tensor defined by the principal values $g_{x} = 2.006$, $g_{y} = 2.003$, and $g_{z} = 2.000$, with Gaussian linewidths of [0.35]~mT. The corresponding zero-field splitting parameters are $D = 5 \times 10^{-4}$~cm$^{-1}$ and $E = 1 \times 10^{-4}$~cm$^{-1}$. This component accounts for approximately 80\% of the total spectrum, primarily the central peak, while the remaining 20\% arises from a secondary defect-related component. The latter is best described by a three-boron centre (TBC)\cite{andrei1976point,moore1972electron,ito2020plane} associated with a nitrogen vacancy ($V_{\mathrm{N}}$), incorporating hyperfine coupling to three equivalent $^{11}$B nuclei with principal values $A_{xx} = A_{yy} = 5~\mathrm{MHz}$ and $A_{zz} = 33~\mathrm{MHz}$ as shown in Figure 5(b). The second case was satisfactorily reproduced using the same set of parameters described above, with the spin system modified to $S = 1/2$. The simulation employed anisotropic $g$-strain values of [0.0004 0.0004 0.0010] and $H$-strain values of [8 8 25] mT, while the zero-field splitting parameters ($D$ and $E$) were omitted. SQUID magnetometry measurements were also performed to probe the magnetic behaviour of the irradiated system. The temperature-dependent magnetic susceptibility (see Figure S7(b), Supplementary Information) exhibits a nonlinear increase at low temperatures, suggesting weak ferromagnetic interactions among the defect centres (see Section S9, Supplementary Information for details). Since first-principles DFT calculations, discussed later in this section, predict a zero-field splitting parameter within the experimentally observed range, and the corresponding simulation provides the best fit with the experimental spectrum, we consider the $S = 1$ fit to represent the most plausible description of the observed EPR signal.

The EPR spectrum recorded at $100~\mathrm{K}$ exhibits features similar to those observed at room temperature and can be reproduced using the same spin-Hamiltonian parameters, indicating that the defect structure remains unchanged over this temperature range. A small field correction of approximately $0.4~\mathrm{mT}$ was applied to account for minor calibration and alignment differences. Angle-dependent EPR measurements (see Figure S8, Supplementary Information) reveal anisotropic behaviour, with a more pronounced rotational dependence in the XY plane at $100~\mathrm{K}$ and only minor variations in the YZ plane. Although the angular dependence could not be quantitatively modelled due to implantation-induced disorder and the presence of multiple defect environments, it qualitatively supports the presence of anisotropic irradiation-induced defect centres\cite{gottscholl2020initialization}(see Section S9, Supplementary Information for details).

The zero-field splitting (ZFS) parameters obtained from the EasySpin simulations, $D = 5 \times 10^{-4}$~cm$^{-1}$ and $E = 1 \times 10^{-4}$~cm$^{-1}$, are exceptionally small compared to those typically reported for localised triplet defects in wide-bandgap materials\cite{gottscholl2020initialization}. Since the dipolar spin--spin interaction scales as $D \propto r^{-3}$ \cite{doi:https://doi.org/10.1002/3527601678.ch34}, this small $D$ value indicates a spatial separation between the two spin-$\frac{1}{2}$ centres forming the triplet state. This is consistent with a weakly coupled defect-pair configuration rather than a single localised defect, and can be viewed within the recently proposed donor--acceptor pair (DAP) framework in hBN\cite{tan2022donor,li2025quantum,robertson2025charge}.We therefore attribute the observed $S = 1$ EPR signal to a spatially separated defect pair, one constituent of which is identified as $V_\mathrm{N}$ through the weakly resolved $^{11}$B hyperfine interaction as shown in Figure 5(c). Motivated by previous studies implicating carbon impurities in optically active defect pairs,\cite{robertson2025charge} and by the fact that carbon is commonly incorporated into hBN during crystal growth and synthesis \cite{onodera2019carbon}, providing a realistic source of substitutional carbon defects, we considered the possibility that irradiation-generated nitrogen vacancies may interact with pre-existing carbon impurities to form $V_\mathrm{N}$--$C_\mathrm{B}$ and $V_\mathrm{N}$--$C_\mathrm{N}$ complexes with an inter-defect separation of approximately 1.0~nm. First-principles calculations were carried out to examine the electronic and magnetic properties of these defect configurations and to evaluate whether they provide viable evidence of the donor--acceptor pair framework.

\begin{figure*}[t]
\centering
\includegraphics[width=\linewidth]{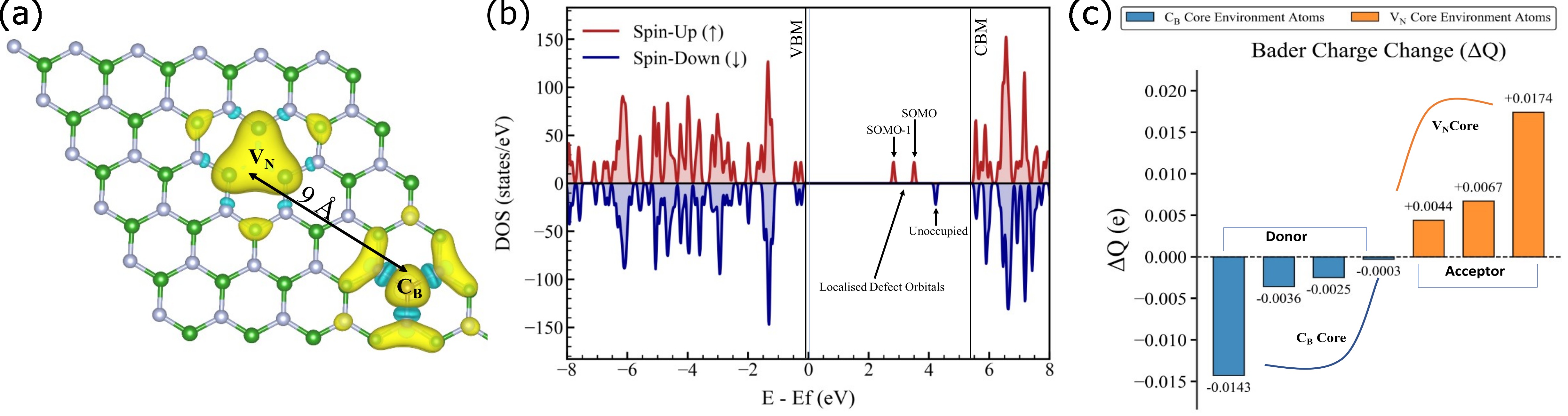}
\caption{\justifying
\textbf{}
(a) Calculated spin density isosurface of the $V_{\text{N}}$--$C_{\text{B}}$ complex in hexagonal boron nitride (hBN). The yellow and cyan contours represent the positive and negative spin densities, respectively, illustrating the localised spin networks distributed at both the nitrogen vacancy ($V_{\text{N}}$) and carbon-at-boron ($C_{\text{B}}$) defect sites.
(b) Spin-resolved total density of states (TDOS) of the $V_{\text{N}}$–$C_{\text{B}}$ defect in hBN. The upper red and lower blue curves correspond to the spin-up ($\uparrow$) and spin-down ($\downarrow$) channels, respectively, with the energy scale plotted relative to the Fermi level.
(c) Quantitative Bader charge change ($\Delta Q$) for the core environment atoms of the $V_{\text{N}}$–$C_{\text{B}}$ defect in hBN.}
\label{fig:DFT_defect_energy_spin}
\end{figure*}

The creation of vacancies in hBN via Kr$^+$ ion bombardment is governed by a delicate balance of local electronic configurations and short-range nuclear repulsion. Previous ion-implantation studies in 2D materials have demonstrated that chemical interactions between the projectile and lattice atoms significantly influence defect production and cannot be captured solely within simple binary-collision models\cite{kretschmer2024atomistic}.To investigate the energetic landscape governing defect formation, the potential energy surface (PES) was scanned 
adiabatically \cite{caro2017stopping} as a function of the Kr projectile trajectory along the $z$-axis as shown in Figure 6(a). As illustrated in the reaction profiles 
Figure 6(b), the interaction energy maximum for boron vacancy formation reaches approximately 731.4~kcal~mol$^{-1}$ at 1.2~\AA, whereas the corresponding maximum for nitrogen vacancy formation is significantly lower, peaking at 0.7~\AA\ with an energy of approximately 249.8~kcal~mol$^{-1}$. This pronounced asymmetry reflects the role of 
local electronic structure and self-consistent charge redistribution  within the hBN lattice during Kr$^+$ approach, and cannot be fully 
captured by purely repulsive pairwise potentials such as the 
Ziegler--Biersack--Littmark (ZBL) model \cite{ziegler1985stopping,krasheninnikov2010ion} (see Section S10, Supplementary Information for a detailed discussion of the interaction regimes and the $V_{\mathrm{N}}$ vs $V_{\mathrm{B}}$ pathways.) Importantly, the observed hierarchy $E^{\mathrm{(N)}}_{\mathrm{max}} < E^{\mathrm{(B)}}_{\mathrm{max}}$ closely follows the established displacement-threshold ordering $T_{d,\mathrm{N}} < T_{d,\mathrm{B}}$ 
reported from first-principles knock-on simulations of hBN \cite{kotakoski2010electron} 
suggesting that the static DFT interaction landscape captures the 
underlying electronic-structure asymmetry that favours nitrogen removal 
relative to boron under irradiation.

The combined EPR measurements and first-principles calculations indicate that the observed paramagnetic centre possesses a triplet ground state ($S = 1$). Previous theoretical studies have shown that isolated nitrogen vacancies in hBN do not support an $S = 1$ ground state, thereby excluding $V_{\mathrm{N}}^{+}$, $V_{\mathrm{N}}^{0}$, and $V_{\mathrm{N}}^{-}$ as the origin of the observed signal\cite{huang2012defect,sajid2018defect}. Consistent with these reports, our singlet--triplet energetic analysis of $V_{\mathrm{N}}^{-}$ (see Section S11, Supplementary Information for details) likewise indicates that the isolated defect cannot account for the experimentally observed triplet resonance.

Spin-polarised DFT calculations were performed on $V_{\mathrm{N}}$--$C_{\mathrm{B}}$ and $V_{\mathrm{N}}$--$C_{\mathrm{N}}$ configurations with an inter-defect separation of $\sim 9.0$~\AA\ and $\sim 11.0$~\AA\ (comparative data for alternative configurations are provided in Section S11 of the Supplementary Information) to study the origin of the proposed defect-pair centre. Among the configurations considered, only the $V_{\mathrm{N}}$--$C_{\mathrm{B}}$ complex stabilises a triplet ground state ($S = 1$), with the corresponding singlet state lying higher in energy by $\Delta E = 23.41~\mathrm{kJ~mol^{-1}}$ as shown in Figure 6(c). In contrast, the $V_{\mathrm{N}}$--$C_{\mathrm{N}}$ configuration remains in a closed-shell singlet ground state, ruling it out as a candidate for the experimentally observed EPR-active centre. The DFT-calculated spin--spin dipolar zero-field splitting parameters as shown in Table 1, namely the axial component $D_{\mathrm{SS}}$ and the rhombic component $E_{\mathrm{SS}}$, which respectively quantify the axial and non-axial anisotropic parts of the magnetic dipole--dipole interaction between unpaired spins, show good agreement with the experimentally fitted values within the typical uncertainty associated with DFT-based ZFS (zero-field splitting) predictions. This consistency supports the validity of the proposed weakly coupled defect-pair model.

\begin{table}[htbp]
\centering
\caption{Calculated ZFS parameters ($D$ and $E$) for $V_{\mathrm{N}}$--$C_{\mathrm{B}}$ defect configuration.}
\label{tab:ZFS_DFT}

\renewcommand{\arraystretch}{1.4} %

\begin{tabular}{|c|c|c|}
\hline
$|D|$ Value (DFT) &
$E$ Value (DFT) &
$V_{\mathrm{N}}$--$C_{\mathrm{B}}$ Separation \\
\hline
$56\times10^{-4}\ \mathrm{cm}^{-1}$ &
$3\times10^{-4}\ \mathrm{cm}^{-1}$ &
$9\ \mathrm{\AA}$ \\
\hline
$20\times10^{-4}\ \mathrm{cm}^{-1}$ &
$5\times10^{-5}\ \mathrm{cm}^{-1}$ &
$11.5\ \mathrm{\AA}$ \\
\hline
\end{tabular}

\end{table}
The spin-density distribution as shwon in Figure 7(a) and singly occupied molecular orbital (SOMO) analysis of the $V_{\mathrm{N}}$--$C_{\mathrm{B}}$ defect configuration reveal two spatially separated spin centres. One spin density is predominantly localised at the nitrogen vacancy, exhibiting the characteristic three-lobe dangling-bond character distributed over the neighbouring boron atoms. A second spin density is localised around the $C_{\mathrm{B}}$ defect and its adjacent nitrogen atoms. While the two spin centres remain spatially distinct, strong spin-polarisation is observed as per the McConnell mechanism, aided by the material's aromaticity, which promotes magnetic exchange between the two sites despite a relatively large distance. Earlier work suggests that such spin polarisation (see Figure S10, Supplementary Information) in strongly aromatic systems promotes stronger exchange, supporting our observations ~\cite{ruiz2006magnetic}.

The total density of states (TDOS), as shown in Figure 7(b), indicates the presence of defect-induced states within the band gap. Two distinct localised in-gap states are observed, corresponding to the singly occupied molecular orbital SOMO and SOMO$-1$, which originate from spin-polarised defect orbitals associated with the unpaired electrons. These states are well isolated from both the valence and conduction band edges, highlighting their deep in-gap character and strongly localised nature, which governs the open-shell electronic structure of the system.

This picture is further supported by the real-space wavefunction (isosurface) analysis (see Figure S10, Supplementary Information), which provides direct visualisation of the spatial distribution of these defect states. The SOMO$-1$ is predominantly localised around the nitrogen vacancy (V${\mathrm{N}}$) site, while the SOMO is mainly centred on the carbon-related defect site (C${\mathrm{B}}$). Although each orbital retains a distinct defect-centred character, a finite spatial overlap is observed between them, indicative of weak electronic coupling between the two defect centres. The resulting spin density distribution is therefore spatially separated yet not completely isolated, consistent with the formation of a weakly coupled $V_{\text{N}}$–$C_{\text{B}}$ donor--acceptor pair.

The Bader Charge analysis \cite{tang2009grid} in Figure 7(c) was performed to quantify the donor-acceptor character within the defect pair. The charge redistribution upon formation of the complex was evaluated as
\[
\Delta Q = Q_{(V_{\mathrm{N}}-C_{\mathrm{B}})} - Q_{\mathrm{isolated}},
\]
where $Q_{(V_{\mathrm{N}}-C_{\mathrm{B}})}$ and $Q_{\mathrm{isolated}}$ denote the Bader charges in the combined and isolated defect configurations, respectively. The analysis reveals a systematic depletion of electron density around the $C_{\mathrm{B}}$ site ($\Delta Q < 0$), accompanied by a corresponding accumulation at the $V_{\mathrm{N}}$ centre ($\Delta Q > 0$). This spatially resolved charge transfer indicates finite electronic coupling between the two defects and is consistent with a donor--acceptor-type interaction within the defect-pair framework.

The weakly coupled, spatially separated spin densities together with the observed charge redistribution are consistent with a donor–acceptor pair (DAP) description of the V$_\mathrm{N}$–C$_\mathrm{B}$ defect configuration. The calculated properties further show good agreement with the experimentally observed $^{11}$B hyperfine interaction and the small zero-field splitting (ZFS) parameter, supporting this configuration as a plausible origin of the room-temperature triplet EPR signal in Kr$^+$-irradiated hBN. Overall, the combined experimental and theoretical results suggest the formation of a spatially separated V$_\mathrm{N}$–C$_\mathrm{B}$ donor–acceptor pair complex defect in Kr$^+$-irradiated hBN \cite{sajid2018defect,toledo2018electron,wu2017first}.

\section{Conclusion}

In conclusion, we demonstrate Kr$^{+}$ ion implantation as a chemically inert, controllable, and scalable route for engineering near-infrared luminescent defects in hBN. SRIM simulations indicate that 40 keV Kr$^{+}$ ions generate a vacancy-rich region confined within $\sim 30.5$ nm of the surface through nuclear-stopping-dominated collisions. Raman spectroscopy confirms preservation of the hBN lattice while revealing irradiation-induced disorder, and photoluminescence measurements show the emergence of a stable defect emission band centred at $\sim 830$ nm that increases with ion fluence. EPR measurements combined with first-principles calculations reveal a paramagnetic centre with $g \approx 2.003$, attributed to a Kr$^{+}$-irradiation-induced $V_{\mathrm{N}}$--$C_{\mathrm{B}}$ donor--acceptor pair defect.

These findings establish Kr$^{+}$ implantation as a practical route for generating stable near-infrared spin defects in hBN and provide a detailed characterisation of the defect configurations responsible for their optical and magnetic signatures. The ability to generate reproducible EPR-active defect ensembles via a single-step, annealing-free ion-implantation process provides a foundation for future work targeting the single-emitter regime and integration with photonic cavity structures.\cite{shaik2021optical,boretti2025advancing,ccakan2025quantum}

\section{Methods}

\subsection{SRIM Simulation}
Monte Carlo simulations were performed using the SRIM-2008 package\cite{ziegler2010srim} to investigate ion stopping, displacement events, and vacancy generation in hBN. The full-cascade damage mode was employed to explicitly account for recoil-induced atomic displacements, which are important for heavy-ion irradiation such as Kr$^{+}$. A target density of 2.1~g~cm$^{-3}$ and a B:N stoichiometric ratio of 1:1 were used. Each simulation comprised $10^{4}$ incident ions. The simulations predicted a mean projected range of $\sim 30.5$~nm and a maximum vacancy-generation within the hBN layer at 40~keV, which guided the selection of the formation.

\subsection{Sample Preparation}

High-quality single-crystal hBN (NIMS, Japan) was mechanically exfoliated from bulk crystals onto quartz substrates using Scotch tape to obtain mechanically exfoliated hBN flakes. Prior to exfoliation, substrates were sequentially cleaned with deionised water, acetone, and isopropyl alcohol (Empura, 99.5\% purity), dried under a nitrogen gas, and treated with UV-ozone (Ossila) for 10~min to remove residual organic contaminants.

\subsection{Atomic Force Microscopy}

AFM measurements were performed in ambient conditions using the Oxford Instruments Asylum Research MFP-3D BIO AFM system at the Department of Biosciences and Bioengineering, IIT Bombay. The surface morphology and thickness of the hBN flakes were characterised in non-contact tapping mode and analysed using Asylum Research Software. The thicknesses of the flakes ranged from $\sim$50 to 250~nm (see Section S4 of the supplementary section).

\subsection{Krypton Ion Implantation}
Ion irradiation was performed using an indigenously developed facility at the Department of Physics, Savitribai Phule Pune University, Pune, India. Krypton ions (Kr$^{+}$) generated from a Penning Ionisation Gauge (PIG)-type ion source were accelerated to 40~keV and implanted into hBN flakes on quartz substrates at room temperature. The implantation chamber operated at a base pressure of approximately $8\times10^{-7}~\mathrm{mbar}$, while the working pressure was maintained between $9\times10^{-6}$ and $3.6\times10^{-5}~\mathrm{mbar}$ depending on the required beam current. Ion fluences ranging from $1\times10^{11}$ to $1\times10^{15}~\mathrm{ions\,cm^{-2}}$ were employed. Beam currents of approximately $0.5~\mu\mathrm{A}$, $0.3~\mu\mathrm{A}$, and $50~\mathrm{nA}$ were used for fluences of $10^{14}$--$10^{15}$, $10^{13}$, and $10^{11}$--$10^{12}~\mathrm{ions\,cm^{-2}}$, respectively, corresponding to exposure times of approximately $560$, $56$, $12$, $7.5$, and $0.75~\mathrm{s}$ for fluences of $10^{15}$, $10^{14}$, $10^{13}$, $10^{12}$, and $10^{11}~\mathrm{ions\,cm^{-2}}$, respectively.

\subsection{Raman and Photoluminescence Spectroscopy}
Raman and photoluminescence (PL) measurements were performed using a Renishaw inVia confocal Raman microscope equipped with a Renishaw Centrus detector (spectral range 200-1064~nm) at the Centre for Research in Nanotechnology and Science (CRNTS), IIT Bombay. Prior to all measurements, the instrument was calibrated using the first-order Raman mode of silicon at 520.4~cm$^{-1}$. Both pristine and Kr$^{+}$-irradiated hBN flakes were characterised under ambient conditions using a 532~nm excitation laser and a 100X objective lens (NA = 0.85).

In Raman spectroscopy, a 2400~lines\,mm$^{-1}$ grating was used to achieve high spectral resolution. The spectra were acquired with an exposure time of 5~s and an incident laser power of approximately 11.1~mW (high power was used to increase the intensity of the inherently weak Raman signal). 

PL measurements were performed using the same optical setup with a 600~lines\,mm$^{-1}$ grating to cover a broader spectral range. The emitted signal was detected using a Renishaw Centrus detector with a detection range of 200--1064~nm. PL spectra were recorded with an exposure time of 1 s and an excitation power of approximately 3.0 mW. 

All PL, Raman spectra presented for comparison were recorded under identical excitation power and exposure time.

\subsection{Low-temperature Photoluminescence measurements}

Temperature-dependent PL measurements were performed using a Lakeshore closed-cycle cryostat at the Centre for Research in Nanotechnology and Science (CRNTS), IIT Bombay. The sample was mounted on the cryostat cold finger using GE Varnish and cooled using Helium exchange gas. The cryostat system was evacuated using a turbomolecular pump to achieve high vacuum ($10^{-5}~\mathrm{mbar}$) prior to cooling. The sample stage temperature was monitored and controlled using the Lakeshore temperature controller, spanning from room temperature to cryogenic temperatures. PL spectra were collected at 15 temperatures from 20 K to 300 K in steps of ~20 K. PL spectra were collected under optical excitation using the same confocal spectroscopy setup described previously at an incident power of 3.83 mW. 

\subsection{Time-resolved Photoluminescence}

Time-resolved photoluminescence (TRPL) measurements were performed using a MicroTime~200 fluorescence lifetime microscope (PicoQuant) equipped with a time-correlated single-photon counting (TCSPC) module at the Centre for Sophisticated Instruments and Facilities (CSIF), IIT Bombay. The defect emission was excited using a pulsed $532~\mathrm{nm}$ laser source. The emitted PL was collected through the microscope objective and directed to the TCSPC detection system. A $532~\mathrm{nm}$ long-pass optical filter was placed in the detection path to stop the excitation laser and allow the total PL from the sample. 

\subsection{Electron Paramagnetic Resonance (EPR)}
Continuous-wave (CW) X-band electron paramagnetic resonance (EPR) measurements were performed using a Bruker EMXplus spectrometer at the Department of Chemistry, IIT Bombay. The spectra were acquired at a microwave frequency of $\sim 9.386$ GHz with a modulation frequency of $100$ kHz, modulation amplitude of $8$ G, and microwave power of $2$ mW. The magnetic field was swept from $200$--$6200$ G with a sweep width of $6000$ G and a spectral resolution of $7500$ points. Measurements were carried out at room temperature and $100$ K to investigate irradiation-induced paramagnetic defect centres in the Kr$^{+}$-implanted hBN sample.

\subsection{Computational Methods}

Electronic structure calculations and geometric relaxations were performed using 
Density Functional Theory (DFT) as implemented in the Vienna Ab initio Simulation 
Package (VASP Version 6.4.3) \cite{kresse1996efficiency}, employing the projector augmented-wave (PAW) method \cite{blochl1994projector}. The PBE-D3 exchange--correlation functional, incorporating Grimme's dispersion correction \cite{grimme2010consistent,perdew1996generalized}, was employed for structural optimisations to account for van der Waals interactions, while single-point energy evaluations were carried out using the HSE06 \cite{heyd2003hybrid,krukau2006influence} hybrid functional. The interaction energy profiles associated with $V_{\mathrm{N}}$ and $V_{\mathrm{B}}$ formation pathways were obtained from an adiabatically constrained PES scan, using a $6 \times 6 \times 1$ hBN supercell containing a single Kr atom positioned along 
selected approach trajectories, to compare the relative short-range interaction energies associated with each displacement pathway. The $6 \times 6 \times 1$ hBN supercell was constructed with an optimised interlayer 
spacing of 3.5~\AA. To eliminate spurious interactions between periodically repeated slabs, a vacuum region of 30~\AA\ was introduced along the out-of-plane ($z$) direction, together with a one-dimensional dipole correction applied along the slab normal. The plane-wave basis set employed a kinetic-energy cutoff of 520~eV. Brillouin zone integrations were performed using a $\Gamma$-centred $2 \times 2 \times 1$ Monkhorst--Pack $k$-point mesh \cite{monkhorst1976special}, corresponding to a reciprocal-space sampling density of approximately 0.03~\AA$^{-1}$. Crystal symmetry operations were explicitly disabled by setting \texttt{ISYM = 0}. All structures were fully relaxed until the residual Hellmann--Feynman forces on each atom were below $10^{-3}$~eV/\AA. Subsequently, the singlet and triplet electronic configurations were fixed and isolated to 
determine the relative energetics of the ground state.

The zero-field splitting (ZFS) parameters and the fine-structure $D$-tensor components of the triplet states were determined by molecular cluster models extracted from the optimised periodic $6 \times 6 \times 1$ and $8\times 8 \times 1$ hBN. These clusters are centred directly on the local defect environment, with peripheral dangling bonds saturated using hydrogen atoms to ensure proper valency and electronic closure. Single-point quantum chemical evaluations were carried out using the ORCA quantum chemistry package (Version 6.1.0) \cite{neese2020orca}. Spin-unrestricted density functional theory (UKS) calculations were employed, utilising the hybrid B3LYP \cite{becke1993density,lee1988development} functional paired with a high-quality def2-TZVP basis set \cite{weigend2005balanced}, supported by the AutoAux auxiliary basis set \cite{stoychev2017automatic} to facilitate efficient resolution-of-the-identity (RI) approximations.
\section*{Author Contributions}

I.S. led the project, performed SRIM simulations, AFM characterisation, Raman and photoluminescence measurements, temperature-dependent spectroscopy, data analysis and visualisation, and prepared the manuscript. R.S. performed the density functional theory calculations and analysed the computational results. M.A. contributed to data analysis and scientific discussions. A.M.S. carried out the Kr$^{+}$ ion implantation experiments and contributed to discussions on ion-solid interactions and defect formation.  M.S. conducted the EPR measurements and contributed to data analysis and interpretation. A.M. performed the time-resolved photoluminescence, contributed to data acquisition, and contributed to scientific discussions. E. assisted with data acquisition, data plotting, and contributed to scientific discussions. J.J.K. contributed to sample fabrication and scientific discussions. A. Khaire assisted with the ion-implantation experiments. K.W. and T.T. provided the hBN crystals. S.S.D. supervised the ion implantation studies, contributed to data interpretation, and reviewed the manuscript. G.R. supervised the theoretical investigations, contributed to the interpretation of the EPR and DFT results, and reviewed and edited the manuscript. A.K. supervised the project, guided the experimental direction of work, secured funding, and critically revised the manuscript.

All authors discussed the results, reviewed the manuscript, and approved the final version.

\section*{Conflict of Interest}
One or more authors of this manuscript are inventors on an Indian patent application related to the work reported here (Indian Patent Application No. 202621071884, filed on 10 June 2026). The patent application was filed by I. Shyam and co-inventors. The remaining authors declare no competing interests.

\section*{Supporting Information}
Supplementary data from this experiment are available.

\section*{Data Availability Request}
All data is available upon reasonable request.

\section*{Acknowledgements}

All the authors acknowledge funding support from the National Quantum Mission, Department of Science and Technology, Government of India. We also acknowledge funding from ANRF via grant number SPR/2023/000175. We thank the Centre for Research in Nanotechnology and Science (CRNTS), IIT Bombay, for access to the Renishaw Raman spectroscopy facility, and Savitribai Phule Pune University, Pune, India, for the ion implantation facility. I.S., R.S, A.M, E, J.J.K. gratefully acknowledge fellowship support from IIT Bombay. M.A. acknowledges the IPDF fellowship from IIT Bombay. M.S. acknowledges the Prime Minister’s Research Fellowship. I.S. thanks Mr Srivatsa Murali (Sreenidhi Institute of Science and Technology) for assistance in preparing the schematic illustration shown in Figure~1(a).

\bibliographystyle{unsrt}   %
\bibliography{References}

\end{document}